\documentclass[12pt]{article}
\usepackage{epsfig, epsf, graphicx}
\usepackage{pstricks, pst-node, psfrag}
\usepackage{amssymb,amsmath,bm}
\usepackage{verbatim,enumerate}
\usepackage{rotating, lscape,natbib}
\usepackage{setspace}
\usepackage{soul}
\usepackage{tikz}
\usetikzlibrary{arrows,calc,tikzmark}
\usepackage{hyperref}
\usepackage{float}
\usepackage{caption}
\usepackage{subcaption}
\usepackage{animate}
\usepackage{mathtools}
\usepackage{multimedia}
\usepackage{comment}
\usepackage{url}
\usepackage[utf8]{inputenc}
\usepackage[english]{babel}

\usepackage{natbib}

\usepackage{color}
\usepackage{verbatim}

\def\boxit#1{\vbox{\hrule\hbox{\vrule\kern6pt
          \vbox{\kern6pt#1\kern6pt}\kern6pt\vrule}\hrule}}

\def\bse{\begin{eqnarray*}}
\def\ese{\end{eqnarray*}}
\def\be{\begin{eqnarray}}
\def\ee{\end{eqnarray}}
\def\bq{\begin{equation}}
\def\eq{\end{equation}}
\def\bse{\begin{eqnarray*}}
\def\ese{\end{eqnarray*}}

\DeclareMathOperator*{\argmin}{arg\,min}
\DeclareMathOperator*{\argmax}{arg\,max}

\begin{document}

\thispagestyle{empty} \baselineskip=28pt \vskip 5mm
\begin{center} {\Huge{\bf Neural Networks for Parameter Estimation in Intractable Models}}
	
\end{center}

\baselineskip=12pt \vskip 10mm

\begin{center}\large
Amanda~Lenzi\footnote[1]{
\baselineskip=10pt Mathematics and Computer Science Division,
Argonne National Laboratory, Lemont, IL, USA.},
Julie Bessac$^{1}$,
Johann Rudi$^{1}$
and~Michael L. Stein\footnote[2]{
\baselineskip=10pt Department of Statistics, Rutgers
University, Piscataway, NJ, USA.}\\
\end{center}

\baselineskip=17pt \vskip 10mm \centerline{\today} \vskip 15mm

\begin{center}
{\large{\bf Abstract}}
\end{center}

We propose to use deep learning to estimate parameters in statistical models when standard likelihood estimation methods are computationally infeasible. We show how to estimate parameters from max-stable processes, where inference is exceptionally challenging even with small datasets but simulation is straightforward. We use data from model simulations as input and train deep neural networks to learn statistical parameters. Our neural-network-based method provides a competitive alternative to current approaches, as demonstrated by considerable accuracy and computational time improvements. It serves as a proof of concept for deep learning in statistical parameter estimation and can be extended to other estimation problems.

\baselineskip=14pt

\par\vfill\noindent
\par\medskip\noindent
{\bf Short title}: Deep learning for parameter estimation

\clearpage\pagebreak\newpage \pagenumbering{arabic}
\baselineskip=26pt

\section{Introduction}\label{sec:intro}

The large datasets that are increasingly available because of advances in data storage and sensor technology inevitably display complex dependencies, bringing  new opportunities and new challenges to statistical modeling and prediction. 
Environmental processes give an important example for which datasets can have complicated interactions across multiple scales of spatial and temporal variability. Specifically, inference for parametric statistical models designed for non-Gaussian data that account for spatiotemporal dependencies are computationally challenging. For instance, consider models for multivariate extremes, where data are usually sampled at a large number of spatial locations but large computational complexities limit their usefulness. In order to circumvent this bottleneck, model approximations that avoid full likelihood computation have been developed. One such approximation is based on composite likelihood, usually constructed from pairs of observations \citep{tawn1988bivariate, padoan2010likelihood, ribatet2012bayesian}. However, even though the maximum pairwise likelihood estimator is strongly consistent and asymptotically normal under mild regularity conditions \citep{xu2011robustness}, no obvious strategy exists for selecting the composite likelihood terms optimally, and the loss in efficiency makes inference for high-dimensional data still challenging \citep{varin2011overview, huser2016likelihood}.

As an alternative to composite likelihood inference in models for spatial extremes,
\citet{stephenson2005exploiting} suggested augmenting the componentwise block maxima data with their occurrence times and showed that partitioning the locations based on whether maxima coincided simplifies the likelihood function substantially. To avoid the bias that emerges from fixing these partitions,
\citet{huser2019full} constructed a stochastic expectation-maximization algorithm for exact maximum likelihood estimation of max-stable processes by treating the partitions as latent variables. Similarly, in a Bayesian setting,
\citet{thibaud2016bayesian} developed a Markov chain Monte Carlo algorithm for max-stable processes conditioning on the partition values. A possible problem with these approaches is that they assume independence in time, and how much bias this temporal dependence might cause is unclear. Moreover, likelihood-based techniques are still hindered by the computational bottleneck of evaluating multivariate max-stable cumulative distribution functions. A key aspect of models for dependence is that often simulation is fast and tractable, although likelihood computation is burdensome. Taking advantage of that, \citet{erhardt2012approximate} employed a simulation-based approach using approximate Bayesian computation (ABC) for estimating parameters of max-stable models. In its most basic form, ABC first samples a set of parameters from the prior distribution, which are then used to simulate data under the assumed statistical model. For each new dataset, a procedure of evaluating discrepancies between the simulations and the observed data must be repeated. As a result, neither of the previously mentioned approaches are well suited for large datasets.
Therefore, developing new inference methods for computationally challenging likelihoods remains an open challenge, as is the case of max-stable processes and many other applications.

In this work we propose an approach to estimate the parameters of intractable statistical models using deep neural networks (NNs). Deep learning algorithms utilize NNs as multiple interconnected layers to predict or classify complex maps between datasets, and in recent years they have shown remarkable success at predicting high-dimensional nonlinear processes. Literature on deep NN techniques applied directly in the context of parameter estimation is scarce. Within the ABC framework, \citet{creel2017neural} proposed to use informative statistics to train a deep NN upon which classical or Bayesian indirect inference may be based. \citet{jiang2017learning} predicted summary statistics from simulated data, which were then used as an estimate of the posterior mean of an ABC procedure.  Recently, \citet{radev2020bayesflow} proposed to perform Bayesian inference using informative summary statistics from the data as the input to invertible NNs. However, these papers provide no general guidance on how to construct suitable summary statistics for the networks in practice. Convolutional neural networks (CNNs) have been successfully applied to represent covariance structures of Gaussian processes by using sensor measurements \citep{liu2018deep}, time-series data \citep{cremanns2017deep}, and spatial data \citep{gerber2020fast}. \citet{rudi2020parameter} proposed to solve a deterministic inverse problem using dense and convolutional layers in NNs, where parameters from a nonlinear system of ordinary differential equations are estimated from simulated data. To our knowledge, the present work is the first study in which deep NN methodologies are utilized with the goal of performing inference based solely on simulated data and parameters. 

Similar to ABC, we perform statistical inference based on simulations rather than direct evaluations of computationally expensive or intractable likelihood functions. A crucial difference from ABC is that instead of measuring similarity as a function of informative summary statistics of the simulated data with the actual observations, our approach learns to perform the mapping directly between data and parameters. Therefore, our method can be seen as a form of automatic ABC, where the data compression is done inside the NN without the difficulties of picking potentially highly sensitive metrics. Our new construction uses simulated data from the statistical model as the input to a deep NN. The output consists of the parameters used to simulate the input data. We leverage the feasibility to simulate arbitrarily large amounts of training data in order to ensure that the deep NNs approximate the actual parameters as well as possible. After the deep NN is trained, the observed data are used as the input in the testing data, returning the model parameters of interest as output.  We show that the deep NN can estimate parameters of max-stable models for spatial extremes with higher accuracy than traditional (approximate) likelihood approaches can, with a considerable speed-up in computations. As the number of parameters increases, a substantial decrease in the computational efficiency is observed for ABC methods, and the same will likely occur with our proposed framework. This work holds promising results for parameter estimation and serves as a proof of concept on how machine learning can be used for statistical inference when the traditional techniques fail.

We demonstrate our framework on a problem of determining parameters from max-stable models. Spatial data in extremes can be viewed as a particular case of 2D images, making NNs with convolutional layers a promising candidate due to their success in learning features of complex images. In this work we initially carry out simulated examples for Brown–-Resnick \citep{kabluchko2009stationary} and Schalther's  \citep{schlather2002models} models and compare the accuracy of our approach with the standard pairwise likelihood technique. We show that CNN architectures are fast to fit and can provide less biased and less variable predictions than the typical approaches provide. Then, using 10-day blocks of reanalysis temperature data in the midwestern United States,  we illustrate the performance and benefits of the proposed framework in terms of accuracy and computational savings over the standard approach.  Another advantage of our approach over pairwise likelihood estimation methods is that no detailed knowledge in the field (e.g., extremes) is necessary since the target statistical computation is done inside the deep NN. Therefore, our deep NN-based framework does not require any assumptions about the data generation process. Instead, it is designed to have a flexible and computationally appealing form, with the network's weights learned by training, which allows accurate representation of a large class of functions. 
Although our focus is on spatial extreme modeling, the new approach can be used for other inference problems as long as simulations from the probability model are possible.

The remainder of this paper is organized as follows. Section~\ref{sec:methods} provides an overview of the proposed approach and briefly introduces CNNs. In Section~\ref{sec:simu} we recall the definition of max-stable processes and inference based on pairwise likelihoods and conduct a simulation study to assess the performance of the estimators. In Section~\ref{sec:applic} we illustrate our method on a dataset of temperature minima, and we conclude with a discussion in Section~\ref{sec:disc}.

\section{Methodology} \label{sec:methods}

This section outlines our new framework for estimating parameters in statistical models by using deep NNs and gives a brief overview of the type of networks used in our examples.

\subsection{Parameter estimation framework} \label{sec:parm-methods}

In what follows, we will use the terminology \textit{parameters} to refer to the parameters in statistical models of interest; the term \textit{NN weights} denotes the parameters of NNs. 
As in classical statistical inference, our goal is to use observed data to infer the underlying characteristics of a stochastic process by estimating parameters of an assumed probability model. Suppose 
$Y$ is a random variable with values in a measurable space and probability density function $p$ with respect to a reference measure $\mu$ (often the Lebesgue measure).
Consider that the probability density function $p(\cdot; \boldsymbol{\theta})$ depends on a finite-dimensional parameter set $\boldsymbol{\theta} \in \Theta \subset \mathbb{R}^K$, where $\Theta$ denotes all possible values of $\boldsymbol{\theta}$ and $K\in\mathbb{N}$. The statistical problem is to recover the unknown $\boldsymbol{\theta}$ given $\mathbf{y} = (y_1, \ldots, y_n)^{\top}$, where $y_i$ are independent copies from $p(\cdot; \boldsymbol{\theta})$.  Evaluating  $p(\cdot; \boldsymbol{\theta})$ at observed data samples $\mathbf{y}$ gives a real-valued function $p(\mathbf{y}; \boldsymbol{\theta})$, which is called the likelihood function. In most cases, this likelihood function is explicitly known and can be evaluated either analytically or numerically for pairs $(\mathbf{y}; \boldsymbol{\theta})$. To estimate $\boldsymbol{\theta}$, one commonly uses the principle of maximum likelihood, which assumes that the most reasonable values are those for which the density of the observed sample is largest. The value $\hat{\boldsymbol{\theta}} = {\hat{\boldsymbol{\theta}}}(\mathbf {y})$ that maximizes the log-likelihood function is called the maximum likelihood estimate (MLE):
\begin{equation}
 \hat{\boldsymbol{\theta}} =  \argmax_{\boldsymbol{\theta} \in \Theta} p(\mathbf{y}; \boldsymbol{\theta}).
 \label{eq:param-mle}
\end{equation}
Often, however, no closed-form solution to the maximization problem is known or available, and the MLE can be found only via numerical optimization. In even more extreme cases, such as many non-Gaussian models for dependencies, writing down the full likelihood is impossible, and advanced numerical optimization strategies are impractical even at moderate dimensions.

We propose to tackle this problem using deep NNs, a type of machine learning that exploits a connected multilayer set of models to learn complex input-output maps. The idea is to use samples of parameters and data pairs to learn the function $p(\cdot; \boldsymbol{\theta})$. In other words, our goal is to find a useful approximation to $p(\cdot; \boldsymbol{\theta})$ that underlies the predictive relationship between $\boldsymbol{\theta}$ and $\mathbf{y}$. Our estimation technique 
is a deep NN taking $\mathbf{y}$ as the independent input data and $\boldsymbol{\theta}$ as the dependent output data
\begin{equation}
    \mathcal{F} : \mathbf{y}\mapsto \boldsymbol{\theta} ; \quad
        \boldsymbol{\hat{\theta}} = \argmin d\{\mathbf{y} , \mathcal{F}_{\mathbf{w}}(\boldsymbol{\theta})\},
    \label{eq:param-nn}
\end{equation}
where $\mathcal{F}_{\mathbf{w}}$ is a function that depends on a set of $l_1$ deep NN weights and $l_2$ NN biases, in other words, $\mathbf{w} = (w_1, \ldots w_{l_1}, b_1, \ldots, b_{l_2})^{\top}$. These weights are determined by training, whereas $d$ in \eqref{eq:param-nn} represents a measure of distance between $\mathbf{y}$ and $\mathcal{F}_{\mathbf{w}}(\boldsymbol{\theta})$. The function $\mathcal{F}_{\mathbf{w}}$ is analogous to the $\argmax$ operation, but it approximates the relation between $\mathbf{y}$ and $\boldsymbol{\theta}$ by means of a deep NN instead of maximizing the likelihood function. In our approach we seek the function $\mathcal{F}$  for predicting $\boldsymbol{\theta}$ given values of the input $\mathbf{y}$. This problem requires a criterion for choosing $\mathcal{F}$, namely, a loss function that penalizes errors in prediction. For regression problems, the mean squared error (MSE) loss is the most common and convenient choice  \citep{friedman2001elements} and is given by
\begin{equation}
    \mbox{MSE}({\mathbf{w}}) = \mathbb{E} \{\boldsymbol{\theta} - \mathcal{F}_{\mathbf{w}}(\boldsymbol{\theta})\}^2.
    \label{eq:mse}
\end{equation}
An optimization algorithm uses the gradients, usually calculated via automatic differentiation \citep{paszke2017automatic}, of the loss function in \eqref{eq:mse} applied to the output of the deep NN and data. 
The idea is that the deep NN finds the set of weights $\mathbf{w}$ such that the fit gets as close to the observed points as possible based on realizations of the output, where proximity is measured by the MSE(${\mathbf{w}}$). For a linear model, a simple closed-form solution to the minimization problem of \eqref{eq:mse} exists; otherwise, the solution often requires iterative methods. Several numerical optimizers built around batch gradient descent methods are implemented to perform this task (e.g., see \citet{kingma2014adam}).

The problem of optimizing $\mathcal{F}_{\mathbf{w}}$ involves generating training data for the deep NN. For this purpose we produce training samples of parameters and data pairs $(\boldsymbol{\theta}^{\mbox{\scriptsize{train}}}_j, \mathbf{y}_j^{\mbox{\scriptsize{train}}})^J_{j=1}$, with $\boldsymbol{\theta}^{\mbox{\scriptsize{train}}}_j = (\theta_{1, j}, \ldots \theta_{K, j})^{\top}$ and $\mathbf{Y}^{\mbox{\scriptsize{train}}}_j = (y_{1, j}, \ldots y_{n, j})^{\top}$, where it is crucial that $\boldsymbol{\theta}^{\mbox{\scriptsize{train}}}_j$ corresponds to configurations covering the parameter domain of interest $\Theta$.
Different approaches can be used to find such configurations. One possible method is to obtain crude estimates of $\boldsymbol{\theta}$ from the observed data $\mathbf{y}$ (e.g., using approximate maximum likelihood methods) and simulate ($\boldsymbol{\theta}^{\mbox{\scriptsize{train}}}_j)^J_{j=1}$ in a large enough neighborhood of those. Then, after the parameter space $\Theta$ has been defined and samples of $(\boldsymbol{\theta}^{\mbox{\scriptsize{train}}}_j)^J_{j=1}$ have been generated, these can be used to simulate corresponding data samples $(\mathbf{y}_j^{\mbox{\scriptsize{train}}})^J_{j=1}$. The number of training samples depends on the difficulty of the problem and directly affects the quality of the estimator \citep{juba2019precision}. However, we find that appropriate transformations of the input and output greatly reduce the number of samples needed for decent accuracy. Once the deep NN has been trained, it can be used to accomplish our goal of performing statistical inference, namely, returning the probability model's 
parameters based on new observed data.

\subsection{Introduction to convolutional neural networks} \label{sec:cnn}

Here we give details on the specific form assumed for $\mathcal{F}_{\mathbf{w}}$ in \eqref{eq:param-nn} responsible for learning the mapping between simulated data (input) and the statistical model parameters (output) of the deep NN. The choice of the deep neural network will likely depend on the type of input and output. In the following sections we analyze spatial 2D data; therefore, we focus on CNNs, a class of deep NNs most commonly applied to visual imagery that accounts for local dependencies in signals. We consider $\mathcal{F}_{\mathbf{w}}$ to be a composition of $m$ nonlinear functions such that $\mathcal{F}_{\mathbf{w}} (\boldsymbol{\theta}) = (f_m \circ f_{m-1} \circ \ldots \circ f_1) (\boldsymbol{\theta})$, where each $f(\cdot)$ is a function of at least one hidden layer. The general structure of a CNN alternates convolution layers followed by pooling layers, with the last layers being fully connected (dense). The convolution operation enables performing weighted averaging of inputs such that the network learns filters that are activated when specific spatial patterns in the input are detected. Formally, given input images $y$ and a kernel $k$, the discrete convolution operator is given by 
\begin{equation}
    K[s_1, s_2] \ast y[s_1, s_2] = \sum_{i= -\infty}^{\infty} \sum_{j= -\infty}^{\infty} K[i, j ]y[s_1 + i, s_2 + j ],
\end{equation}
where $s_1$ and $s_2$ are pixels of the image. The sums are finite in practice since the kernel $K$ has compact support and depends on the number of pixels. Each kernel $K$ is a matrix of trainable weights (also known as filters) applied to small regions of the input image, 
and the same kernel is translated across the dimensions of the image. Several independently trained kernels can be applied when multiple filters are used in a convolutional layer. Depending on the values of the kernel weights, one can get different properties associated with the image, such as edges of objects. We refer to \citet{pinaya2020convolutional} for more information on CNNs and to \url{https://tensorflow.org/api_docs} for instructions on the TensorFlow implementation of convolutional layers.

\section{Proof of concept for max-stable processes } \label{sec:simu}

Max-stable distributions are commonly used for studying extreme events recorded in space and time. As a proof of concept, we exemplify the methodology from Section~\ref{sec:methods} by estimating parameters of max-stable processes. These processes are well known for having full likelihoods that are computationally intractable even in moderate dimensions. However, simulation is manageable. With this example, we aim to demonstrate that our method can accurately recover the parameters of a model with intractable likelihood by learning the mapping from raw data. We use here the algorithm provided by \citet{schlather2002models} for simulation of max-stable models implemented in the \texttt{SpatialExtremes} \texttt{R}-package.

In Section~\ref{sec:max-stab}
we briefly outline max-stable distributions and processes and the most commonly used approach for estimating parameters in such models. 
Sections~\ref{sec:simu-setup} and \ref{cnn-arch} outline the  parameter estimation setup and the CNN used in the estimation procedure, respectively.
In Sections~\ref{sec:max-br} and \ref{sec:max-pexp}
the performance of the CNN estimator is assessed and compared with a benchmark through a simulation study using two popular max-stable models. 

\subsection{Definition and the pairwise likelihood approach} \label{sec:max-stab}

Consider a sequence of independent and identically distributed (i.i.d.) stochastic processes $Y_1(\mathbf{s}), Y_2(\mathbf{s}), \ldots$, where $\mathbf{s} \in \mathcal{S} \subset \mathbb{R}^d$ is a spatial site. A max-stable process Z is the limit process of normalized pointwise block-maximum process (with block size $m$), provided that sequences of functions $a_m(\mathbf{s}) > 0$ and $b_m(\mathbf{s})$ exist:
\[
Z(\mathbf{s}) = \lim_{m\to\infty} a_m(\mathbf{s})^{-1} \left[\max_{m}\{Y_1(\mathbf{s}), \ldots, Y_m(\mathbf{s})\} - b_m(\mathbf{s})\right], \quad \mathbf{s} \in \mathbb{R}^d,
\]
in other words, that pointwise maxima of independent copies of $Z(\mathbf{s})$ remain in the same family up to a location and scale functions. 
Each marginal distribution of $Z(\mathbf{s})$ is a generalized extreme value (GEV) distribution with location, scale, and shape parameters that potentially depend on the spatial locations \citep{de2007extreme}. 

We now introduce a representation that will serve as the basis for constructing a parametric max-stable process in the following sections. Let $\{\xi_i\}_{i\geq 1}$ be the points of a Poisson process on $(0,\infty)$ with intensity $d \Lambda (\xi) = \xi^{-2}d\xi$, and let $W_1(\mathbf{s})$,  $W_2(\mathbf{s})$, $\ldots$ be independent copies of a nonnegative stochastic process such that $W(\mathbf{s}) \geq 0$ with mean equal to one. The processes $W_i$ and the points of the Poisson process $\{\xi_i\}_{i\geq 1}$ are assumed to be independent. Then
\begin{equation}
    Z(\mathbf{s}) = \max\limits_{i \geq 1} \xi_i W_i(\mathbf{s}), \quad \mathbf{s} \in \mathcal{S},
    \label{eq:spect}
\end{equation}
is a max-stable process with unit Fréchet margins: $\mbox{P}\{Z(\mathbf{s} \leq z)\} = \mbox{exp}(-1/z)$, $z>0$. Suitable choices for $W(\cdot)$ yield to different max-stable processes \citep{schlather2002models}.
In the next sections we will consider two different choices: Brown-Resnick \citep{kabluchko2009stationary} and Schalther's model \citep{schlather2002models}. 
\begin{itemize}
    \item The Brown–-Resnick model arises when $W_i(\mathbf{s}) = \mbox{exp}\{ \epsilon_i(\mathbf{s}) - \gamma(\mathbf{s})\}$ in \eqref{eq:spect}, where  $\epsilon_i(\mathbf{s})$ are independent copies of a standard stationary Gaussian process such that $\epsilon(\mathbf{0}) = 0$ almost surely and with semivariogram $\gamma(\mathbf{h}) = (\lVert\mathbf{h}\rVert/\lambda)^\nu$, where  $\mathbf{h}$ is the spatial separation, $\lambda > 0$ is the range, and $\nu \in (0, 2]$ is a smoothness parameter. Brown-Resnick processes are popular in practice  because of their flexibility compared with other choices and the ability to generalize several stationary max-stable models, such as the geometric Gaussian model \citep{davison2012statistical}. For a precise justification of its practical performance, see \citet{thibaud2015efficient}.
    \item The second model is the characterization proposed in \citet{schlather2002models}, with $W_i(\mathbf{s})= \sqrt{2\pi} \mbox{max}\{0, \epsilon_i(\mathbf{s})\}$, where $\epsilon_1(\mathbf{s}), \epsilon_2(\mathbf{s}), \ldots$ are independent copies of a stationary Gaussian process with unit variance and correlation function $\rho$ that is a powered exponential: $\rho(\mathbf{h}) = \mbox{exp}\{-(\lVert\mathbf{h}\lVert/\lambda)^\nu\}, \lambda>0, \nu \in (0, 2]$. 
\end{itemize} 

\citet{schlather2002models} showed that one can obtain exact simulations of max-stable models on a finite sampling region if the support of the random Poisson process $\xi$ (see \eqref{eq:spect}) is either included in a ball $b(o; r)$ or is uniformly bounded by a constant C. For Brown-Resnick and Schlather's models, these conditions are not satisfied since Gaussian processes are not uniformly bounded. However,  \citet{schlather2002models} introduced approximations for the constant $C$ such that if $\mbox{P}[\mbox{max}\{0; \xi_i(\mathbf{s})\} > C]$ is sufficiently small, the simulation procedure is still accurate.

It follows from \eqref{eq:spect} that the joint cumulative distribution of $Z(\mathbf{s})$  at a finite collection of sites $\{\mathbf{s_1}, \ldots, \mathbf{s}_D\} \subset S$ is given by
\begin{equation}
    p(Z(\mathbf{s}_1) \leq z_1, \ldots, Z(\mathbf{s}_D) \leq z_D) = \mbox{exp}\{-V( z_1, \ldots,  z_D)\},
    \label{eq:cdf}
\end{equation}
where the exponent function $V( z_1, \ldots,  z_D) = \mbox{E}[\mbox{max}\{W(\mathbf{s}_1)/z_1, \ldots, W(\mathbf{s}_D)/z_D\}]$, satisfies homogeneity and marginal constraints \citep{de1984spectral}. By differentiating the distribution \eqref{eq:cdf} with respect to its variables $z_1 , \dots, z_D$, one can deduce the corresponding density function, or the full likelihood  as
\begin{equation}
    f(z_1, \ldots, z_D) =  \mbox{exp}\{-V( z_1, \ldots,  z_D)\} \sum_{\pi \in \mathcal{P}} \prod_{l=1}^{L} \{-V_{\pi_l}( z_1, \ldots,  z_D)\},
    \label{eq:pdf}
\end{equation}
where $\mathcal{P}$ denotes the collection of all partitions $\pi = \{\pi_1, \ldots,  \pi_L\}$ of $\{1, \ldots, D\}$ and $V_{\pi_l}$ denotes the partial derivative of the function $V$ with respect to the variables indexed by the set $\pi_l$. The sum in Equation \eqref{eq:pdf} is taken over the set of all possible partitions, which equals the \textit{Bell} number of order $D$. This leads to an explosion of terms even for a moderate $D$, and the full likelihood quickly becomes intractable.
\citet{castruccio2016high} concluded that with modern computer technologies, full likelihood inference is still limited to $D \leq 12$. In the simple bivariate case, however, the density is computationally tractable, which explains the common practice of adopting the pairwise likelihood in place of the full likelihood (see, e.g., \citet{padoan2010likelihood, davis2013statistical, shang2015two}). The corresponding log weighted pairwise likelihood for model \eqref{eq:cdf} is given by
\begin{equation}
    l(\boldsymbol{\phi}) =  \sum_{(j_1, j_2) \in \mathcal{P}} \alpha_{j_1, j_2} \Big[ \mbox{log} \{ V_1(z_{j_1}, z_{j_2}) V_2(z_{j_1}, z_{j_2}) - V_{12}(z_{j_1}, z_{j_2})\} - V_1(z_{j_1}, z_{j_2}) \Big ],
    \label{eq:pl}
    \end{equation}
where $\boldsymbol{\phi} \in \boldsymbol{\Phi} \subset \mathbb{R}^{p}$ is the vector of unknown parameters and $\alpha_{j_1, j_2} \geq 0$ denotes the likelihood weight assigned to the pair $\{j_1, j_2\}$. We suppress the dependence of $V_1$, $V_2$, and $V_{12}$ on $\boldsymbol{\phi}$ for notation simplicity. The weights are often set to one only for pairs of locations up to a certain distance: $\mathbf{h} = \lVert \mathbf{s}_1 - \mathbf{s}_2 \lVert < \delta$, where $\delta > 0$ is an appropriate cut-off distance chosen, for instance, via cross-validation \citep{ribatet2008spatialextremes}. The  block maximum recorded at the $j$th station is denoted by $z_j$, where $ \mathcal{P} = \{ (j_1, j_2): 1 \leq j_1 < j_2 \leq D \}$. For multiple independent replicates of the $z_j$'s, the maximum pairwise likelihood estimator $\hat{\boldsymbol{\phi}}$ in \eqref{eq:pl} is strongly consistent and asymptotically Gaussian. The asymptotic variance for such misspecified likelihood estimators is of the sandwich form \citep{padoan2010likelihood}. However, calculating this sandwich variance is time-consuming, since it requires working with four points at a time. The variability of $\hat{\boldsymbol{\phi}}$ may also be assessed by using block bootstrap techniques.

\subsection{Proposed parameter estimate setup} \label{sec:simu-setup}

For simplicity, we describe below our proposed estimation approach for simulated data on uniform margins, such that we estimate only the dependence
parameters. We simulate $I$ independent replicates $(\boldsymbol{\theta}_i, \mathbf{y}_i)^{I}_{i=1}$ (also known as testing data), where $\boldsymbol{\theta}_i = (\lambda^{\mbox{\scriptsize{test}}}_i, \nu^{\mbox{\scriptsize{test}}}_i)^\top$ are the range and smoothness of the max-stable process and $\mathbf{y}_i = \{y_i(\mathbf{s}_1), \ldots, y_i(\mathbf{s}_D) \}^{\top}$ are the corresponding realizations (either Brown–-Resnick or Schalther's). 
We consider here $I=16$ parameter scenarios with values of $\lambda^{\mbox{\scriptsize{test}}}_i$ and $\nu^{\mbox{\scriptsize{test}}}_i$ on a (not necessarily equally spaced) grid. 
For each pair $(\lambda^{\mbox{\scriptsize{test}}}_i, \nu^{\mbox{\scriptsize{test}}}_i)$, max-stable samples are generated at $D = 25^2$ (i.e., $25 \times 25$ images) in $[0, 20]^2$. We then obtain estimates of $\boldsymbol{\theta}_i$ using two different approaches:

\begin{enumerate}
    \item CNN estimator $\hat{\boldsymbol{\theta}}_{i}^{\mbox{\scriptsize{CNN}}} = (\hat{\lambda}_{i}^{\mbox{\scriptsize{CNN}}}, \hat{\nu}_{i}^{\mbox{\scriptsize{CNN}}})^\top$: We simulate training data $\mathbf{y}^{\mbox{\scriptsize{train}}}_j$ each of size $D = 25^2$ from max-stable models with parameters randomly sampled from $\lambda_j^{\mbox{\scriptsize{train}}} \sim \mbox{Unif}(a^{\mbox{\scriptsize{train}}}_{\lambda}, b^{\mbox{\scriptsize{train}}}_{\lambda})$ and $\nu_j^{\mbox{\scriptsize{train}}} \sim \mbox{Unif}(a^{\mbox{\scriptsize{train}}}_{\nu}, b^{\mbox{\scriptsize{train}}}_{\nu})$, for $j = 1, \ldots, 2000$. The choice of $a^{\mbox{\scriptsize{train}}}_{\theta}$ and $b^{\mbox{\scriptsize{train}}}_{\theta}$, for $\theta = \lambda, \nu$, is informed by the testing parameters $\boldsymbol{\theta}_i$ to ensure that the training set covers the region of interest.
    The CNN is trained by using log($\mathbf{y}^{\mbox{\scriptsize{train}}}$) as the input, where $\mathbf{y}^{\mbox{\scriptsize{train}}}$ is a tensor with dimensions $2000 \times 25 \times 25$ and $\Big\{\mbox{log}(\boldsymbol{\lambda}^{\mbox{\scriptsize{train}}}), \mbox{log}\Big(\frac{\boldsymbol{\nu}^{\mbox{\scriptsize{train}}}}{2 - \boldsymbol{\nu}^{\mbox{\scriptsize{train}}}}\Big) \Big\}^{\top}$ as the output with size $2000 \times 2$.
    The logarithm is used here as a variance-stabilizing transformation to help with numerical issues during training, and the denominator corresponding to $\nu$ is chosen so that this parameter is not greater than two.
    Once trained, the CNN is used to return predictions $\hat{\boldsymbol{\theta}}_{i}^{\mbox{\scriptsize{CNN}}}$ based on logarithm-transformed testing images $\mathbf{y}_i$, for $i=1, \ldots, 16$.
    
    \item Pairwise-likelihood estimator $\hat{\boldsymbol{\theta}}_{i}^{\mbox{\scriptsize{PL}}} = (\hat{\lambda}_{i}^{\mbox{\scriptsize{PL}}}, \hat{\nu}_{i}^{\mbox{\scriptsize{PL}}})^\top$: The models are fitted to data by using the pairwise likelihood (see \eqref{eq:pl}) through the \texttt{R}-function \texttt{fitmaxstab}. This function performs optmization using another \texttt{R}-function, \texttt{optim}, and we set the method to L-BFGS-B. After a small cross-validation study, we find that including pairs that are more than three units apart worsens the estimation, besides increasing the fitting time. Therefore, we combine only likelihood contributions from pairs that are at most three units apart with equal weights $\alpha_{i_1, i_2} = 1$, and the remaining weights are set to zero. We initialize the parameters by giving the optimizer multiple random starting pairs around the actual values. Then we run the full optimization from the five pairs with the highest pairwise likelihood among the random pairs as the starting point. We show the results using the pair with the pairwise likelihood maximizer as initial values.
    
\end{enumerate}

\subsection{CNN architecture} \label{cnn-arch}

We use a sequential CNN taking simulated images as input and mapping it onto two scalar values. Table~\ref{tab:cnn-arch} provides a summary of the settings used to train the CNN using simulated data from the Brown-Resnick and Schlather's model. We report the number of layers and how they transform the dimension of the input tensors. For example, the first row of Table~\ref{tab:cnn-arch} shows that the model input is a tensor of shape [–, 25, 25, 128], where \lq–' indicates an
arbitrary number of samples, the number \lq25' relates to the dimension of the images, and \lq128' stands for the number of filters. 
Then, the first convolution layer transforms the input tensor into another tensor of output shape [–, 13, 13, 128] by using 128 filters with a kernel size of $3\times 3$ (see the second row). 

The proposed CNN consists of three pairs of convolutions with varying numbers of filters. We keep three dense layers at the end of the network with 8, 16, and 2 units. The total number of trainable weights of the CNN is 168,558. For all the layers, we use the rectified linear unit (ReLU) activation function, which has been shown to be beneficial for regression type models \citep{hastie2009elements}. The implementation of our algorithms is carried out by  using TensorFlow/Keras built in the \texttt{R}-software.
The CNN weights are initialized randomly; and  to train the network, we employ the widely used Adam optimizer \citep{kingma2014adam} with a learning rate of 0.01.  The training is performed for 32 epochs, where in each epoch the CNN weights are updated utilizing a batch size of 40 samples from the full training dataset, which here consists of $J=2000$ samples.

\begin{table}[tb]
\centering
 \begin{tabular}{ ccccc } 
 \hline
 Layer Type & Output Shape & Filters & Kernel Size &  Parameters  \\ \hline
  2D conv & [-, 25, 25, 128] & 128 & $3 \times 3$ & 1280 \\ 
  2D conv & [-, 13, 13, 128] & 128 & $3 \times 3$ &  147584 \\ 
  2D conv & [-, 7, 7, 16] & 16 & $3 \times 3$ & 18448  \\ 
  dense & [-, 4] & &   & 1028  \\ 
  dense & [-, 8] & &  & 40  \\ 
  dense & [-, 16] &  &  & 144  \\ 
  dense & [-, 2] &  &  & 34  \\ 
 \hline 
 Total trainable weights: & & & & 168,558
\end{tabular}
\caption{Summary of the CNN model. It is a sequential model taking input of shape [–, 25, 25, 128] and mapping it to two scalar values of shape [–, 2].}
\label{tab:cnn-arch}
\end{table}

\subsection{Results for the Brown-Resnick model} \label{sec:max-br}

\begin{figure}[htb!]
	\centering
	\scriptsize{(a)} \\
			\begin{subfigure}[b]{0.63\textwidth}
		\includegraphics[width=1\textwidth]{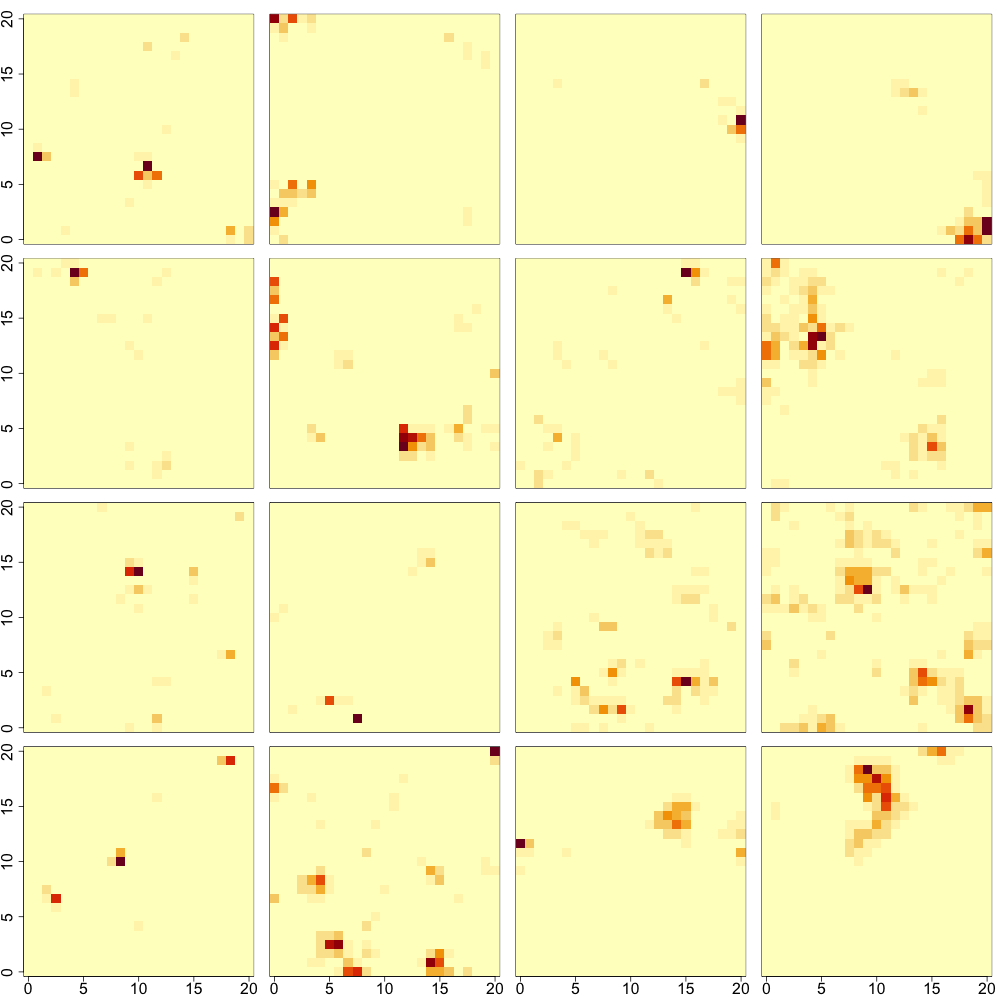}
	\end{subfigure}
	\\
	     \scriptsize{(b)} \\
     \begin{subfigure}[b]{0.63\textwidth}
      \includegraphics[width=1\textwidth]{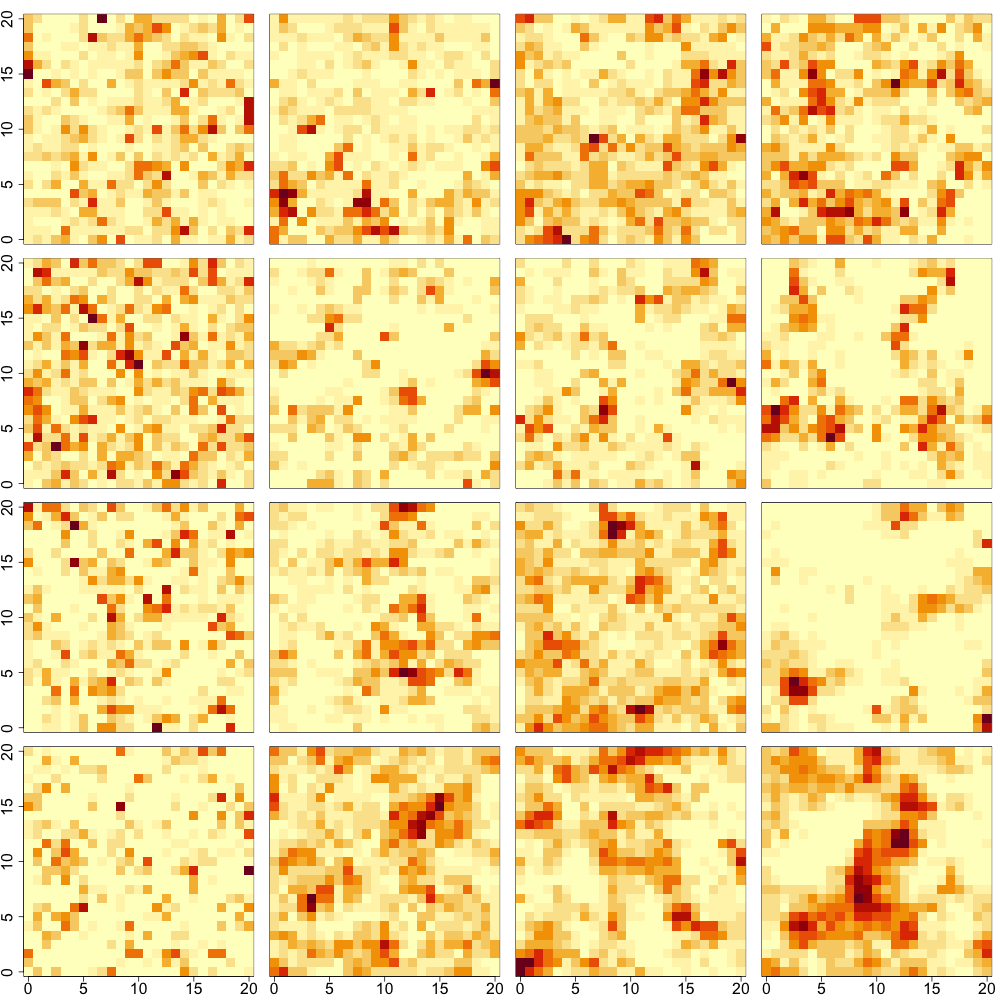}
	\end{subfigure}
	\caption{Simulations from the Brown–Resnick (a) and Schlather's (b) models on  $25 \times 25$ images covering $[0, 20]^2$. Small to large range values ($\lambda$) are shown from left to right and rough to smooth ($\nu$) processes are from top to bottom.
	}
		\label{fig:sim_br_pexp}
\end{figure}

In this section we show a comparison between CNN and pairwise likelihood for estimating the parameters of a Brown-Resnick model. A description of this model is given in Section~\ref{sec:max-stab}, and details of the parameter estimation set up for both methods can be found in Section~\ref{sec:simu-setup}. The testing set comprises 16 pairs of simulated parameters and data, where the parameters are all possible combinations ($\lambda^{\mbox{\scriptsize{test}}}, \nu^{\mbox{\scriptsize{test}}})^{\top}$, with $\lambda^{\mbox{\scriptsize{test}}} \in [0.50, 0.75, 1.00, 1.50]$ and $\nu^{\mbox{\scriptsize{test}}} \in [0.80, 1.05, 1.30, 1.55]$. For each combination, which we call a scenario, we simulate 50 i.i.d. realizations of the respective Brown-Resnick process, which will be used to assess the uncertainty of the estimators.  Examples of realizations covering the simulated parameters are illustrated in Figure~\ref{fig:sim_br_pexp}~(a).


The training set for the CNN has output values drawn from $\lambda_j^{\mbox{\scriptsize{train}}} \sim \mbox{Unif}(0.1, 3)$ and $\nu_j^{\mbox{\scriptsize{train}}} \sim \mbox{Unif}(0.5, 1.9), j=1, \ldots, 2000$, so that it covers the range of the true output by a reasonable margin. Once the network has converged, we use it to predict Brown-Resnick parameters on the test data. Scatterplots of the estimated transformed range (log($\lambda$)) versus transformed smoothness (log\{$\nu/(2-\nu$)\}) (with the number on the axes in the original scale) from the CNN (green) and pairwise likelihood (red) are displayed in Figure~\ref{fig:scatter_transf}~(a). The $\times$ symbol in each plot represent the truth. Each scatterplot comprises 50 independent data replicates that are estimated independently. When fitting the pairwise likelihood, we notice the importance of using the maximizer among multiple starting values (we used 20 values around the true and performed full optimization from the five pairs with the highest pairwise likelihood; see Section~\ref{sec:simu-setup} for details). 
This figure shows that the CNN works consistently well, whereas the pairwise likelihood performance varies between scenarios. As  expected from estimating long-range dependencies, there is an increased variability for larger values of this parameter, especially from the pairwise likelihood estimator. In particular, the pairwise likelihood tends to underestimate the smoothness parameter. This is more evident when the range and smoothness are both large (see bottom right corner of Figure~\ref{fig:scatter_transf}~(a)). For large smoothness, there were a few data replicates for which the pairwise likelihood failed to estimate both parameters (see the last row of Figure~\ref{fig:scatter_transf}~(a)). For rougher fields with medium ranges, there is a large variability in the estimates, in most cases underestimating the smoothness. Based on this study, the CNN results in an overall more minor estimation bias and variance than the pairwise likelihood, with most estimations closer to the actual parameter values.

Table~\ref{tab:cnn_log_tab} reports values of root mean squared error (RMSE), mean absolute error (MAE), and mean bias from predicting Brown-Resnick parameters ($\lambda$; $\nu$) using the CNN and pairwise likelihood approaches. Each score is the result of 50 replicates and the 16 scenarios combined. Results for the Brown-Resnick model (columns 2 and 3) confirm that the CNN estimator outperforms the pairwise likelihood overall. This is more evident for the smoothness parameter, with higher biases and RMSEs and MAEs about 1.6 and 2.7 times larger for the pairwise likelihood. Biases resulting from pairwise likelihood estimates are in accordance with the scatterplots in Figure~\ref{fig:scatter_transf}~(a) and might be explained by the low efficiency of using only pairs of variables. Moreover, we consider one replicate for each test data, and increasing this number could make the estimation more stable for the pairwise. However, investigating higher-order composite likelihood methods is outside this paper's scope, and we refer the reader to \citet{genton2011likelihood} for details on this approach. 

\subsection{Results for Schlather's model} \label{sec:max-pexp}

Similarly to the experiment for the Brown-Resnick model described in the preceding section, we simulate testing sets from Schlather's model comprising $50$ independent copies of $16$ parameter scenarios of size $D = 25^2$ in $[0,20]$. Here, the parameters ($\lambda^{\mbox{\scriptsize{test}}}, \nu^{\mbox{\scriptsize{test}}})^{\top}$ used to simulate data consist of all pair combinations from 
$\lambda^{\mbox{\scriptsize{test}}} \in [0.50, 1.50, 2.0, 2.5]$ and $\nu^{\mbox{\scriptsize{test}}} \in [0.80, 1.05, 1.30, 1.55]$. The dataset to train the CNN contains simulated images of the same size where the range and smoothness are generated according to $\lambda_j^{\mbox{\scriptsize{train}}} \sim \mbox{Unif}(0.1, 3)$ and $\nu_j^{\mbox{\scriptsize{train}}} \sim \mbox{Unif}(0.5, 1.8), j=1, \ldots, 2000$. Figure~\ref{fig:sim_br_pexp}~(b) shows examples of simulations from the Schlather's model. This model produces data that are more noisy and presents a more challenging scenario when compared with the Brown-Resnick process presented in the preceding section (see Figure~\ref{fig:sim_br_pexp}~(a)).

\begin{figure}[htb!]
	\centering
	\scriptsize{(a)} \\
			\begin{subfigure}[b]{0.78\textwidth}
      \includegraphics[width=1\textwidth]{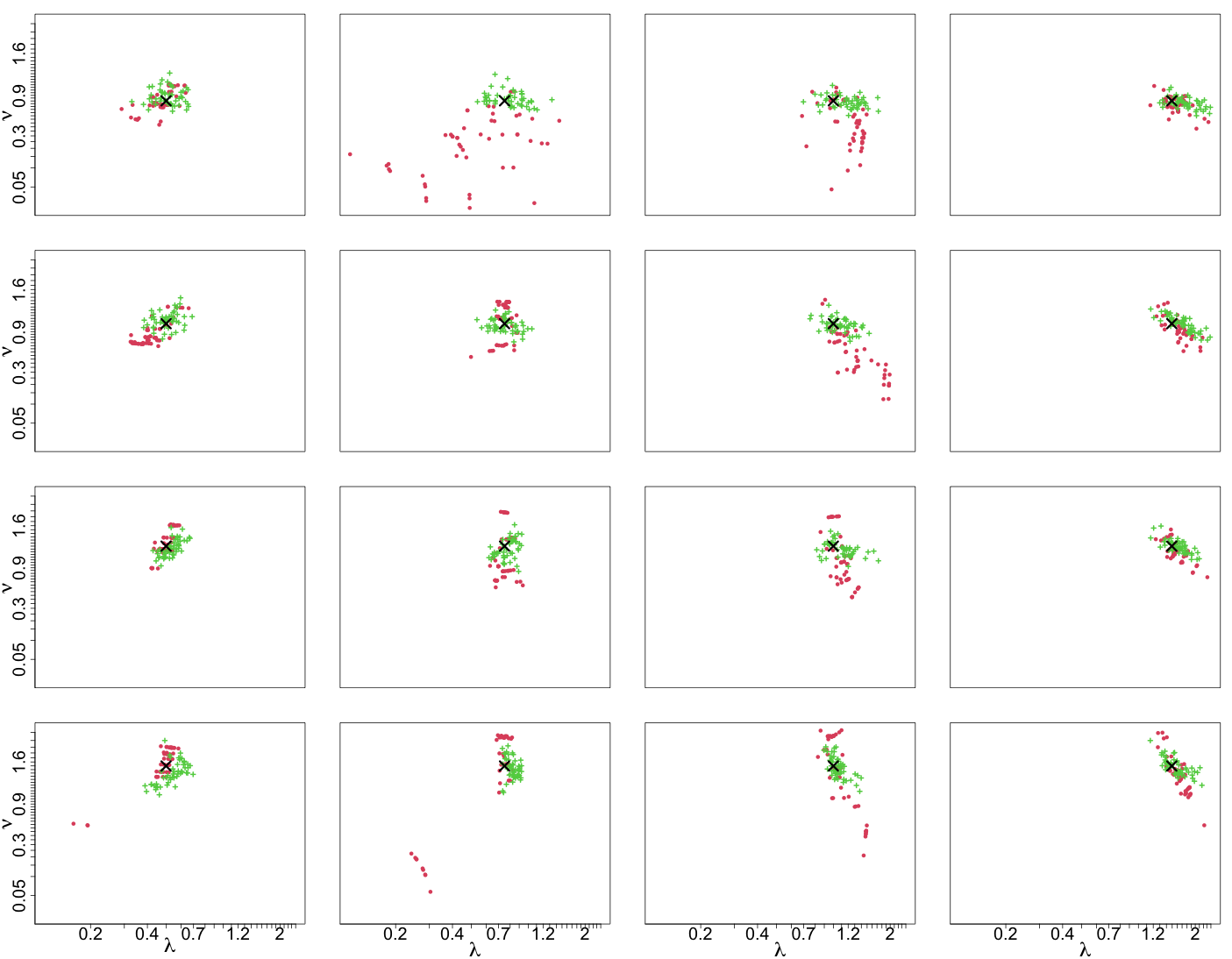}
	\end{subfigure}
	\\
	     \scriptsize{(b)} \\
     \begin{subfigure}[b]{0.78\textwidth}
      \includegraphics[width=1\textwidth]{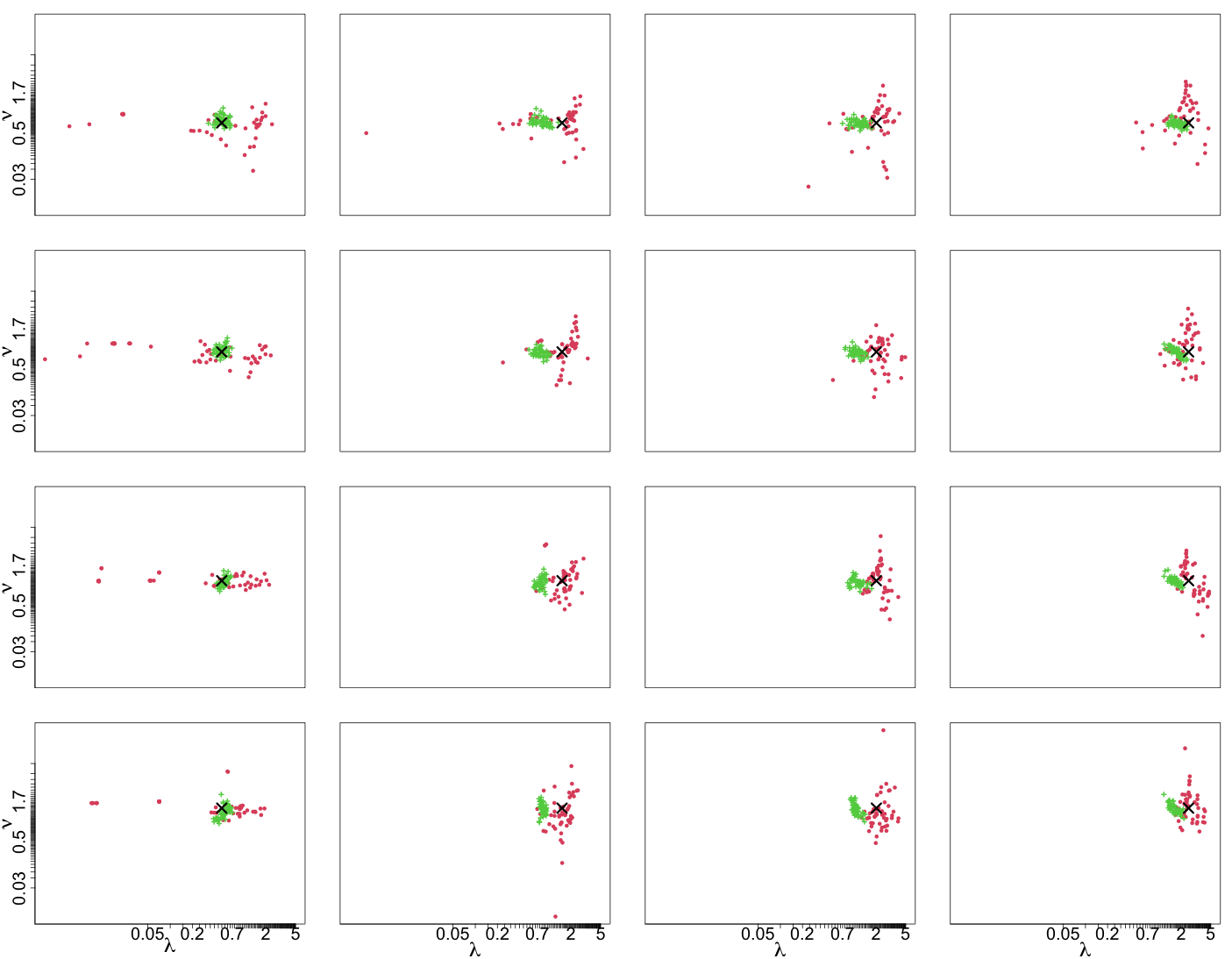}
	\end{subfigure}
	\caption{Scatterplots of estimated parameters on the transformed scales (with numbers on the axes on the untransformed scales). Each plot shows 50 independent estimates from the Brown-Resnick (a) and Schlather's (b) models using the CNN (green) or PL (red) with the first initial values that maximizes the PL (red). Small to large ranges are shown from left to right, and rough to smooth are from top to bottom. The $\times$'s are the true values.
	}
		\label{fig:scatter_transf}
\end{figure}

Each scatterplot in Figure~\ref{fig:scatter_transf}~(b) displays estimated values of log($\lambda$) versus log\{$\nu/(2-\nu$)\} for 50 replicates and a different scenario where the estimation is performed by using either the CNN (green) or pairwise likelihood (red). Compared with the analyses for  Brown-Resnick processes (see Figure~\ref{fig:scatter_transf}~(a)), both methods produce estimates with less variability, and the bias from the pairwise is less prominent here. Exceptions are scenarios with small ranges for which the pairwise likelihood underestimates the range in a few datasets (see the first column of Figure~\ref{fig:scatter_transf}~(b)). The CNN produces small biases when predicting both parameters and all scenarios, and the variability is slightly larger for large-range cases.
The results suggest that the CNN estimator performs well overall and constantly improves the pairwise likelihood results. 
We notice that the accuracy of the CNN is regardless of the range of training values since all the estimates under CNN are never near the boundary of the range of simulated parameter values. Although only small biases arise from the pairwise likelihood in estimating the range for most parameter values, in a couple of instances the method performed poorly, producing estimates about two to three times smaller than the true ones. In contrast, the CNN estimates larger ranges well, and outliers are extremely unlikely to occur. 

The improved pattern from the CNN over the pairwise likelihood estimator is confirmed in Table~\ref{tab:cnn_log_tab}, with RMSE and MAE values that are around $60\%$ to $80\%$ smaller. The only case where the pairwise likelihood beats the CNN is in terms of mean biases for the smoothness parameter, which is primarily seen for small ranges, namely, when the spatial dependence is relatively weak (see Figure~\ref{fig:scatter_transf}~(b)). In such cases the CNN estimates of smoothness seem to have a distribution that is not strongly dependent on the actual smoothness. The computations were carried out on the 8-core processor Intel Core i9 with 2.4 GHz. The elapsed times (in seconds) for fitting and predicting the parameters, excluding the time for producing training samples, from the Brown-Resnick and Schlather’s models from the CNN are 58 and 45, respectively. In contrast, for the pairwise likelihood, these times are 1,305 and 14,960, respectively. The CNN-based parameter estimation, excluding the time for producing training samples, reduced the computational time by a fraction of 300. Overall, similarly to what we observed for the Brown-Resnick model, the CNN provides better results than does the pairwise likelihood, requiring massively less computational resources. The success of the CNN confirms our previous conclusions. It even more strongly supports the need for alternative methods to overcome the difficulty of the suboptimal approaches currently used to fit intractable models. Moreover, such results are expected to improve in higher-dimensional settings, where the loss in efficiency of pairwise likelihood estimators is more significant.


\begin{table}[htb!]
    \centering
\begin{tabular}{ |c|c|c|c|c|c| } 
 \hline
  & \multicolumn{2}{|c|}{Brown-Resnick}  &
  \multicolumn{2}{|c|}{Schlather's} \\ \hline
  & CNN & PL  & CNN & PL  \\ 
    RMSE & \textbf{0.45;0.38} & 0.56;0.62 &  \textbf{0.66;0.49} & 0.83;0.62 \\
     MAE & \textbf{0.14;0.11} & 0.24;0.30 &  \textbf{0.33;0.18} & 0.55;0.31 \\
     Mean Bias & \textbf{0.09;-0.02} & 0.05;-0.15 &  -0.23;\textbf{0.01} & \textbf{0.12};-0.02 \\
 \hline
\end{tabular}
    \caption{RMSE, MAE, and mean bias from the Brown-Resnick and Schlather's models using the CNN and the pairwise likelihood approaches. The two numbers in each column for the first three rows represent scores for estimating range and smoothness, respectively.}
    \label{tab:cnn_log_tab}
\end{table}

\section{Application to temperature data} \label{sec:applic}

In this section we discuss an application of our method to a surface temperature reanalysis dataset, which we describe in Section~\ref{sec:data}. Section~\ref{sec:data-est} outlines the estimation framework and highlights the additional steps needed to estimate parameters from real data compared with the simulation study in Section~\ref{sec:simu}. In Section~\ref{sec:data-results} we analyze the results.

\subsection{NLDAS data and preprocessing} \label{sec:data}

Reanalysis temperature data are extracted from the North American Land Data Assimilation System (NLDAS-2) \citep{mitchell2004multi,xia2012part1, xia2012part2} and are freely available at \url{https://ldas.gsfc.nasa.gov/nldas/v2/models}. 
NLDAS data is quality-controlled and available at a grid spacing of 12 km and the hourly temporal resolution. Hourly data are extracted between 1991 and 2019.  

\begin{figure}[htb!]
 \centering
 \includegraphics[width=0.95\textwidth]{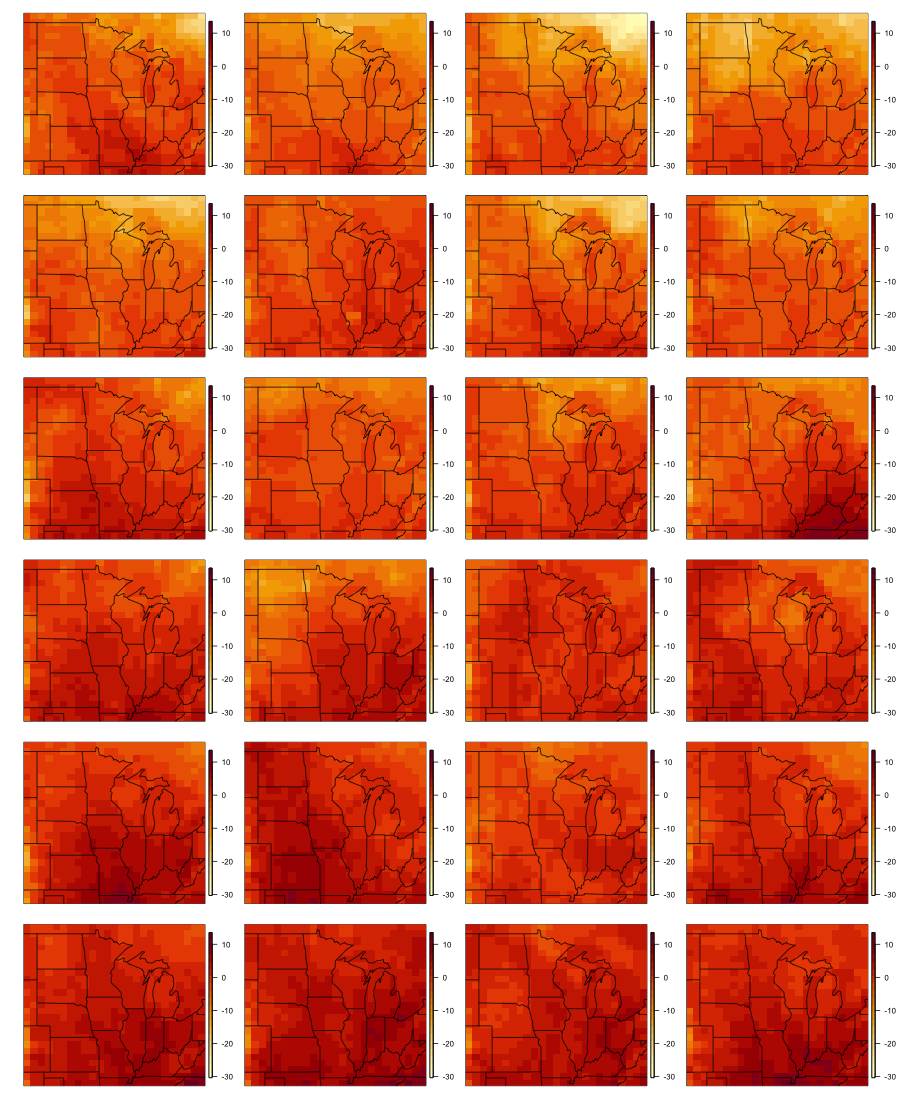}
	    \caption{Temporal evolution of temperature minima from six 10-day data blocks over April-May (rows 1-6) during 1991, 2000, 2009, 2019 (columns 1-4).
	    }
	    	\label{fig:ex_gev}
\end{figure}

We select locations in the midwestern United States on a $107 \times 55$ grid-scale and perform spatial downsampling by selecting 25 equally spaced points in the latitude and longitude directions so that images are the same size as in the simulation study (see Section~\ref{sec:simu}). To decrease the effect of seasonality, we restrict the data to 60 days over April-May, although there is still large variability within this period. We consider daily minima, and the models are fit to the negative of this quantity over six 10-day blocks, resulting in 174 datasets of size $25 \times 25$. Figure~\ref{fig:ex_gev} shows images of minimum temperature data from the six 10-day data blocks over April-May (rows 1-6) at four equally spaced years between the 29 years available (rows 1-4). As expected, temperature minimas are lower at the beginning of April.

\begin{figure}[htb!]
 \centering
 \centering
\includegraphics[width=1\textwidth]{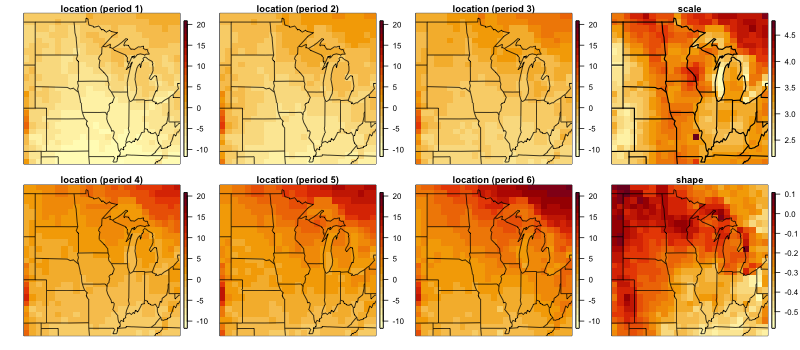}
	    \caption{(a) Estimated location for each of the 6 blocks of data, scale, and shape parameters obtained from the individual fit of the GEV model.
	    }
	    	\label{fig:ex_params_gev}
\end{figure}

Although the spectral representation of max-stable processes in \eqref{eq:spect} relies on Fréchet margins, in practice the observations are rarely a unit Fréchet distribution. We follow the usual approach of fitting a GEV distribution before transforming the data to the unit Fréchet scale. We model marginal distributions with GEVs using componentwise minima over six 10-day blocks at each site separately. We estimate a single scale and shape parameter for the considered April-May period. In contrast, the location parameter is specific to each of the six 10-day blocks to better meet the identically distributed assumption. Quantile-quantile plots (not shown) suggest that a GEV reasonably approximates marginal distributions.  Images of the estimated locations, scale, and shape are displayed in Figure~\ref{fig:ex_params_gev}. Location parameters for each of the 10-day blocks gradually increase as the weather gets warmer. Shape parameters are primarily negative, indicating a bounded upper tail. Negative shape parameter estimates are commonly found when fitting GEV distributions of temperature extremes; see, for example,  
\citet{kharin2000changes, huang2016estimating}. Exceptions are close to the Great Lakes (see mid-upper right corner), where the positive shape parameter implies a heavy tail. We then use these estimated parameters from the GEV to transform the data to the unit Fréchet scale. 

\subsection{Implementation details of the estimation framework} \label{sec:data-est}

\begin{figure}[htb!]
 \centering
		\includegraphics[width=1\textwidth]{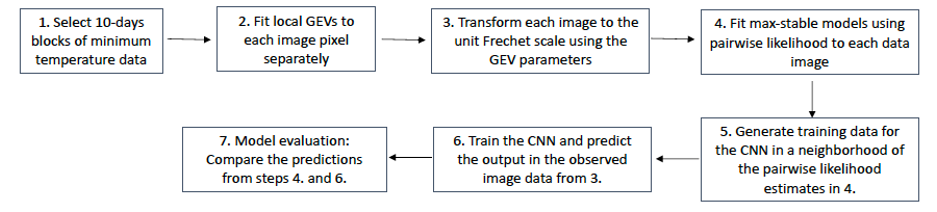}
	\caption{Diagram of the proposed framework for estimating parameters from temperature data. Knowledge about the mapping
between simulated data and parameters is compactly encoded within the weights of the CNN.}
		\label{fig:diag_cnn}
\end{figure}

In Section~\ref{sec:simu} we generated the training set for the CNN  based on the data and parameters from the testing set, which were simulated and were therefore known. In this section the parameters of the testing set are unknown. Consequently, we first need to define a range of parameter values that reflect the assumed max-stable distribution of the unit Fréchet transformed data. The diagram in Figure~\ref{fig:diag_cnn} illustrates the several steps needed to obtain parameter estimates of max-stable processes using the temperature data. Notice that in the simulated setup described in Section~\ref{sec:simu}, only steps 4, 5, 6, and 7 were needed. Different approaches can be taken to choose the region to sample the parameters from. We present one possible method, where the permissible parameter region from fitting a pairwise likelihood to the data is expanded by simulating data in a broader grid with uniform probabilities. 

\begin{figure}[htb!]
 \centering
		\includegraphics[width=0.6\textwidth]{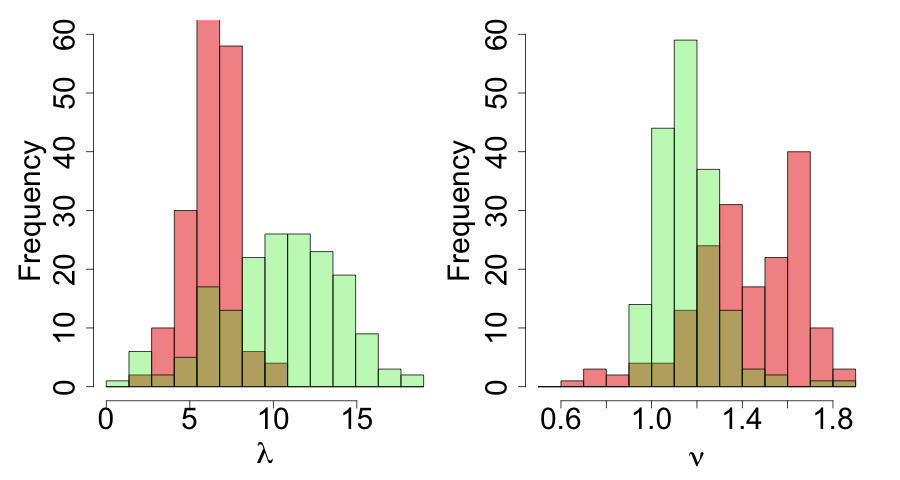}
	\caption{Histograms of the estimates of range (left) and smoothness (right) from a Brown-Resnick model fitted to the Fréchet transformed data by using pairwise likelihood (red) and CNN (green).
	}
		\label{fig:params_wpl_CNN}
\end{figure}

The first task is to fit Brown-Resnick models to each of the 174 transformed data images separately by using the pairwise likelihood estimator \eqref{eq:pl}, for which pairs of locations are at most three units apart. Figure~\ref{fig:params_wpl_CNN} displays histograms of the pairwise likelihood (red) estimated range (left) and smoothness (right). Whereas the range parameter varies from approximately $12\%$  to $50\%$ of the total size of the domain $[0, 20]$, the estimated smoothness is often relatively large. Next, we train the CNN with simulated training data of size 2,000 generated by sampling both parameters independently from uniform distributions. The lower and upper bounds of the uniform are obtained by perturbing the pairwise likelihood estimates based on their  standard deviations: 
 \begin{align}
    \lambda^{\mbox{\scriptsize{train}}}_j &\sim \mbox{Unif}\{\hat{\lambda}_{\mbox{\scriptsize{min}}}^{\mbox{\scriptsize{PL}}} - 3 \times \mbox{sd}(\hat{\boldsymbol{\lambda}}^{\mbox{\scriptsize{PL}}}) \wedge 0,\hat{\lambda}_{\mbox{\scriptsize{max}}}^{\mbox{\scriptsize{PL}}} + 3 \times \mbox{sd}(\hat{\boldsymbol{\lambda}}^{\mbox{\scriptsize{PL}}}) \}, \quad j = 1, \ldots, 2000 \nonumber \\
        \nu_j^{\mbox{\scriptsize{train}}} &\sim \mbox{Unif}\{\hat{\nu}_{\mbox{\scriptsize{min}}}^{\mbox{\scriptsize{PL}}} - 3 \times \mbox{sd}(\hat{{\boldsymbol{\nu}}}^{\mbox{\scriptsize{PL}}}) \wedge 0, \hat{\nu}_{\mbox{\scriptsize{max}}}^{\mbox{\scriptsize{PL}}} + 3 \times \mbox{sd}(\hat{{\boldsymbol{\nu}}}^{\mbox{\scriptsize{PL}}}) \wedge 2 \}, \quad j = 1, \ldots, 2000, \label{eq:train_cnn}
\end{align}   
where $a \wedge b$ denotes the maxmium between $a$ and $b$,
$\hat{{\boldsymbol{\Theta}}}^{\mbox{\scriptsize{PL}}} = (\hat{\Theta}_{1}^{\mbox{\scriptsize{PL}}}, \ldots, \hat{\Theta}_{174}^{\mbox{\scriptsize{PL}}})^{\top}$, $\hat{\Theta}_{\mbox{\scriptsize{min}}}^{\mbox{\scriptsize{PL}}} = \mbox{min}(\hat{\Theta}_{1}^{\mbox{\scriptsize{PL}}}, \ldots, \hat{\Theta}_{174}^{\mbox{\scriptsize{PL}}})$ and $\hat{\Theta}_{\mbox{\scriptsize{max}}}^{\mbox{\scriptsize{PL}}} = \mbox{max}(\hat{\Theta}_{1}^{\mbox{\scriptsize{PL}}}, \ldots,  \hat{\Theta}_{174}^{\mbox{\scriptsize{PL}}})$ for $\Theta = \lambda, \nu$.
Using the range and smoothness values generated from \eqref{eq:train_cnn}, we simulate Brown-Resnick processes with spatial dimension $D=25^2$.
We then train a CNN model using the logarithm transformed data from these simulations as inputs and the variance stabilizing transformation on the outputs (see Section~\ref{cnn-arch} for details). The CNN model configuration is summarized in Table~\ref{tab:cnn-temp}. After the model is trained, we predict the output based on input images of logarithm transformed data. Histograms of estimated parameters at their original scale from the CNN are displayed in Figure~\ref{fig:params_wpl_CNN} (green). Compared with the pairwise likelihood estimates (red), the range parameter (left) estimated from the CNN  is usually larger and more uniform across datasets. The smoothness (right) is closer to normally distributed with a higher probability of values close to one.

\begin{table}[htb!]
\centering
 \begin{tabular}{ cccccc } 
 \hline
 Layer Type & Output Shape & Filters & Kernel Size &  Parameters  \\ \hline
  2D conv & [-, 25, 25, 128] & 128 & $3 \times 3$ & 1280 \\ 
  2D conv & [-, 13, 13, 128] & 128 & $3 \times 3$ & 147584  \\ 
  2D conv & [-, 7, 7, 128] & 16 & $3 \times 3$ & 147584  \\ 
   2D conv & [-, 4, 4, 16] & 16 & $3 \times 3$ & 18448  \\ 
  dense & [-, 4] & &   & 260  \\ 
  dense & [-, 8] & &  & 72  \\ 
  dense & [-, 8] &  &  & 72  \\ 
  dense & [-, 2] &  &  & 18  \\ 
 \hline 
 Total trainable parameters: & & & & 315,358
\end{tabular}
\caption{Summary of the CNN model. It is a sequential CNN taking inputs of shape [–, 25, 25, 1] and mapping it to two scalar values of shape [–, 2]. 
}
\label{tab:cnn-temp}
\end{table}

\subsection{Results} \label{sec:data-results}

We detail here the results from the CNN and pairwise likelihood approaches with the data and implementation described in the preceding sections. For clarity, the results are presented on the Fréchet transformed data scale.

A useful quantity to visualize spatial extremes is the so-called extremal coefficient
$\theta(\mathbf{s}_1; \mathbf{s}_2) \in [1, 2], \mathcal{D} = \{\mathbf{s}_1; \mathbf{s}_2\}$, giving a measure of extremal dependence between two stationary max-stable random fields, $Z(\mathbf{s}_1)$ and $Z(\mathbf{s}_2)$, where  $\theta(\mathbf{s}_1; \mathbf{s}_2) = 1$ corresponds to perfect dependence and $\theta(\mathbf{s}_1; \mathbf{s}_2) = 2$ to independence. The modified version of a variogram, called F-madogram,  proposed by \citet{cooley2006variograms}, is defined as $v_F(\mathbf{s}_1; \mathbf{s}_2) = 0.5 \times \mathbb{E}\{|F[Z(\mathbf{s}_1)] - F[Z(\mathbf{s}_2)]|\}$, where $F(z) = \mbox{exp}(-1/z)$ are unit Fréchet margins. The F-madogram is used here as a summary statistic for the extremal coefficient through the relation $2v_F(\mathbf{s}_1; \mathbf{s}_2) = \frac{\theta(\mathbf{s}_1; \mathbf{s}_2) - 1}{\theta(\mathbf{s}_1; \mathbf{s}_2) + 1}$. 

\begin{figure}[htb!]
 \centering
 \begin{subfigure}[b]{0.4\textwidth}
		\includegraphics[width=1\textwidth]{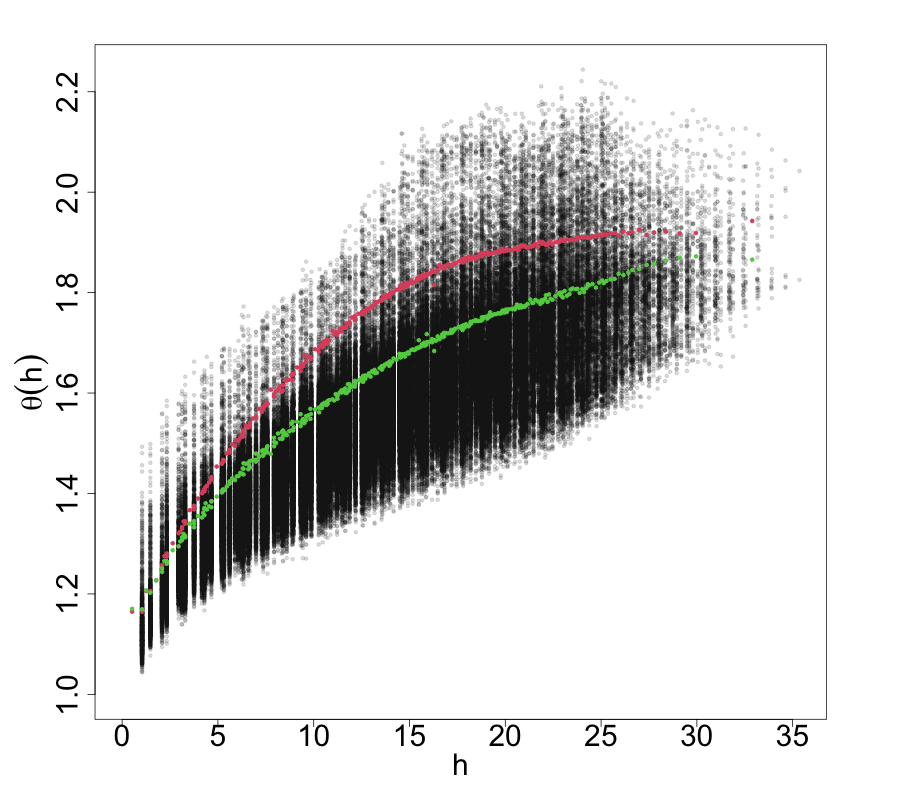}
		\end{subfigure} 
	  \begin{subfigure}[b]{0.4\textwidth}   		
	  \includegraphics[width=1\textwidth]{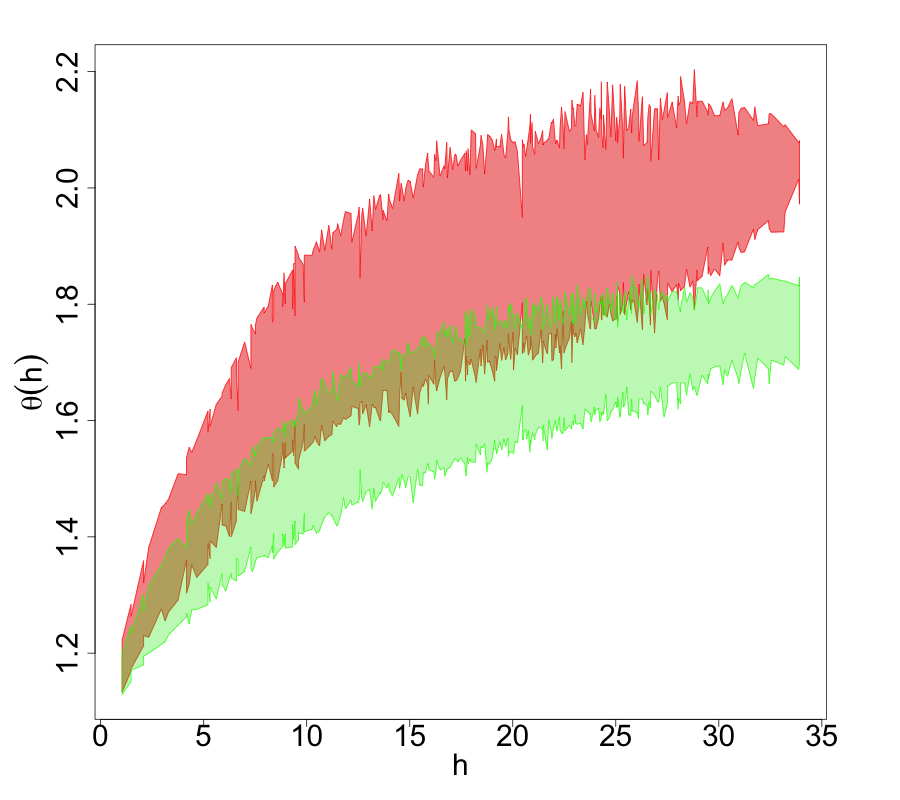}
	  \end{subfigure}
	    \caption{Left: F-madogram estimates for the validation datasets (black points) and the estimated extremal coefficient functions from the pairwise likelihood (red) and CNN (green) using 100 bins. 
	    Right: Example of nonbinned F-madogram estimates using as parameter values the estimates obtained from one of the images in the testing set.
	    }
	    	\label{fig:fmado}
\end{figure}

The left panel of Figure~\ref{fig:fmado} compares the empirical F-madogram estimates (black dots) of the extremal coefficients between pairs of sites, plotted against the spatial distance $\mathbf{h} = \lVert \mathbf{s}_i - \mathbf{s}_j \rVert$, with respect to their pairwise likelihood (red curve) and CNN (green curve) model counterparts. The empirical F-madogram is calculated from 174 datasets at the $25^2$ locations, and estimates of the binned F-madogram for the two models are obtained by using 100 bins.  The pairwise likelihood fit appears to underestimate extremal dependence at distances greater than five (red curve is generally above the back dots). 
In contrast, the CNN provides a more realistic estimation of spatial dependence, matching empirical and estimated extremal coefficients. 

The uncertainty from both model fits can be assessed from the right panels of Figure~\ref{fig:fmado}, which show nonbinned estimated F-madograms of the extremal coefficients based on 200 simulations from Brown-Resnick processes using as parameter values the estimates obtained from one of the images in the testing set. For this example, the CNN (green) estimates a considerable larger range parameter compared with the pairwise likelihood (red) ($\hat{\lambda}_i^{\mbox{\scriptsize{CNN}}} = 28.6$ and $\hat{\lambda}_i^{\mbox{\scriptsize{PL}}} = 8.9$), which better matches the empirical spatial dependence (see black dots in the left panel). Consequently, only the CNN can capture pairs with high extremal dependencies (e.g., $\theta(\mathbf{h}) < 1.5$), a feature that is more evident at larger spatial separation (the green area is below the red in the rightmost plot). 

\begin{figure}[htb!]
 \centering
		\includegraphics[width=0.85\textwidth]{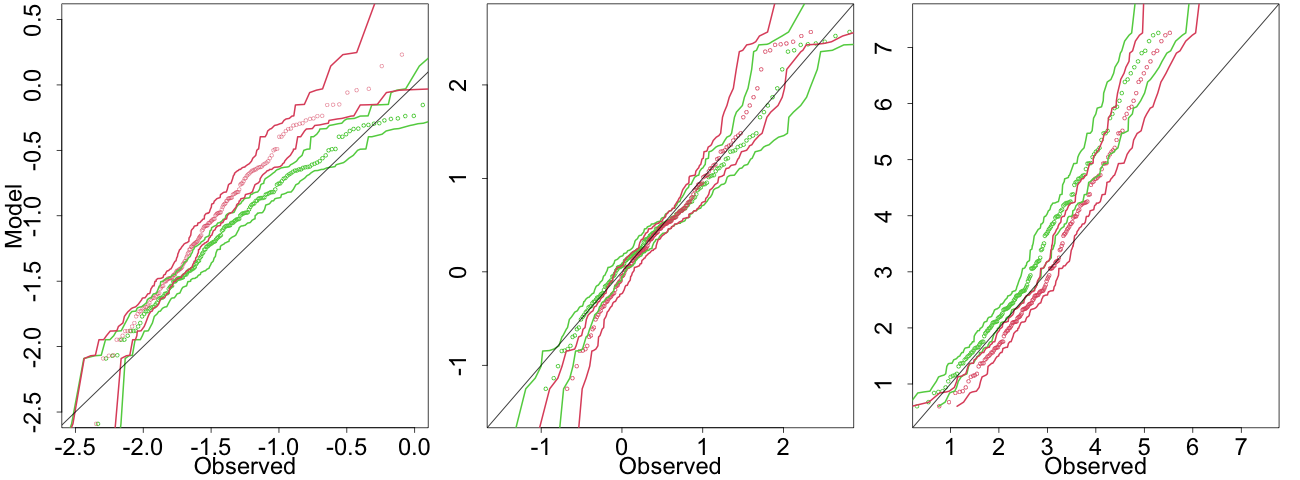}
	    \caption{Comparison of the observed versus predicted minima (left), mean (middle), and maxima (right) with  95\% confidence intervals from the pairwise likelihood (red) and CNN (green) fits. 
	    }
	\label{fig:qqplot_mmm}
\end{figure}

The quantile-quantile plots in Figure~\ref{fig:qqplot_mmm} compare the observed and predicted minima (left), mean (middle), and maxima (right) from the pairwise likelihood (red) and CNN (green) estimates using 200 simulated replicates for each test data and all the $25^2$ sites. Overall $95\%$ confidence intervals calculated by using quantiles from the simulations are also shown with solid lines. The CNN outperforms the pairwise likelihood for the lower quantiles and for predicting minima and mean. For maxima and moderate quantiles, the estimated values are below the diagonal for pairwise likelihood and above for CNN, suggesting that the spatial dependence in the extremes is underestimated by the pairwise likelihood and overestimated by the CNN. More significant maxima quantiles are overestimated by both methods, although the uncertainty is higher since there are fewer cases in the dataset.


\section{Discussion} \label{sec:disc}

In this work we have proposed a new approach to estimate parameters in statistical models based on deep neural networks. As a proof of concept, we have tested the methodology based on the highly challenging problem of inference for max-stable distributions and processes. 
We illustrated the benefit of using deep NN models over the classical pairwise likelihood approach through simulated and real-world data. Results from our simulation study for the popular Brown–Resnick and Schlather's models showed better performances (more negligible bias and variance) of our method compared with the current approach with speedups in computations. Unlike pairwise likelihood, the proposed deep learning approach relies on simulations of the full model and does not involve approximations of spatial dependence. 

We follow the paradigm that if it is possible to simulate from a model, one can perform inference. Our approach uses data from simulations as input to a deep NN and learns to predict statistical parameters. Similar to the well-known ABC, the idea is that the deep NN becomes a stand-in for compressing data and recognizing model parameters. However, a key difference from ABC is that our method automatically maps data and parameters inside the deep NN, eliminating the problem of choosing summary statistics and metrics to compare data. Another difference is that in ABC the entire estimation procedure must be run again from scratch for each individual dataset, whereas in the proposed framework the estimation is performed upfront in a single training phase, followed by a cheaper prediction phase able to handle multiple testing sets. 
Moreover, our new approach benefits from recent advances in algorithms for conventional deep NNs, which can learn features automatically from highly complex images while being fast and easy to apply. Problems with our approach will emerge when there are many parameters in the model, up to a point where generating training data for high-dimensional parameter spaces becomes practically intractable.

In this work we have focused on modeling multivariate extremes and followed a proof of concept approach. We can use the same idea presented here for inferring parameters in other statistical models. In particular, non-Gaussian models for dependencies tend to be hard to estimate by using classical approaches. Extensions for future work include nonhomogeneous Poisson processes \citep{moller2003statistical}, epidemiology models \citep{lawson2008bayesian}, discrete stochastic population dynamics models \citep{wood2010statistical}, and random set models. 

In summary, we show that deep NN models are easy to construct, are fast to fit, and show promising results for performing statistical inference, especially when the usual techniques fail.

\baselineskip=18.5pt

\bibliography{sample}

\begin{thebibliography}{}

\bibitem[Castruccio et~al., 2016]{castruccio2016high}
Castruccio, S., Huser, R., and Genton, M.~G. (2016).
\newblock High-order composite likelihood inference for max-stable
  distributions and processes.
\newblock {\em Journal of Computational and Graphical Statistics},
  25(4):1212--1229.

\bibitem[Cooley et~al., 2006]{cooley2006variograms}
Cooley, D., Naveau, P., and Poncet, P. (2006).
\newblock Variograms for spatial max-stable random fields.
\newblock In {\em Dependence in Probability and Statistics}, pages 373--390.
  Springer.

\bibitem[Creel, 2017]{creel2017neural}
Creel, M. (2017).
\newblock Neural nets for indirect inference.
\newblock {\em Econometrics and Statistics}, 2:36--49.

\bibitem[Cremanns and Roos, 2017]{cremanns2017deep}
Cremanns, K. and Roos, D. (2017).
\newblock Deep {G}aussian covariance network.
\newblock {\em arXiv preprint arXiv:1710.06202}.

\bibitem[Davis et~al., 2013]{davis2013statistical}
Davis, R.~A., Kl{\"u}ppelberg, C., and Steinkohl, C. (2013).
\newblock Statistical inference for max-stable processes in space and time.
\newblock {\em Journal of the Royal Statistical Society: SERIES B: Statistical
  Methodology}, pages 791--819.

\bibitem[Davison et~al., 2012]{davison2012statistical}
Davison, A.~C., Padoan, S.~A., Ribatet, M., et~al. (2012).
\newblock Statistical modeling of spatial extremes.
\newblock {\em Statistical Science}, 27(2):161--186.

\bibitem[De~Haan et~al., 1984]{de1984spectral}
De~Haan, L. et~al. (1984).
\newblock A spectral representation for max-stable processes.
\newblock {\em The Annals of Probability}, 12(4):1194--1204.

\bibitem[De~Haan and Ferreira, 2007]{de2007extreme}
De~Haan, L. and Ferreira, A. (2007).
\newblock {\em Extreme Value Theory: An Introduction}.
\newblock Springer Science \& Business Media.

\bibitem[Erhardt and Smith, 2012]{erhardt2012approximate}
Erhardt, R.~J. and Smith, R.~L. (2012).
\newblock Approximate {B}ayesian computing for spatial extremes.
\newblock {\em Computational Statistics \& Data Analysis}, 56(6):1468--1481.

\bibitem[Friedman et~al., 2001]{friedman2001elements}
Friedman, J., Hastie, T., Tibshirani, R., et~al. (2001).
\newblock {\em The {E}lements of {S}tatistical {L}earning}, volume~1.
\newblock Springer Series in Statistics.

\bibitem[Genton et~al., 2011]{genton2011likelihood}
Genton, M.~G., Ma, Y., and Sang, H. (2011).
\newblock On the likelihood function of {G}aussian max-stable processes.
\newblock {\em Biometrika}, pages 481--488.

\bibitem[Gerber and Nychka, 2020]{gerber2020fast}
Gerber, F. and Nychka, D.~W. (2020).
\newblock Fast covariance parameter estimation of spatial {G}aussian process
  models using neural networks.
\newblock {\em Stat}, page e382.

\bibitem[Hastie et~al., 2009]{hastie2009elements}
Hastie, T., Tibshirani, R., and Friedman, J. (2009).
\newblock {\em The Elements of Statistical Learning: Data Mining, Inference,
  and Prediction}.
\newblock Springer Science \& Business Media.

\bibitem[Huang et~al., 2016]{huang2016estimating}
Huang, W.~K., Stein, M.~L., McInerney, D.~J., Sun, S., and Moyer, E.~J. (2016).
\newblock Estimating changes in temperature extremes from millennial-scale
  climate simulations using generalized extreme value (gev) distributions.
\newblock {\em Advances in Statistical Climatology, Meteorology and
  Oceanography}, 2(1):79--103.

\bibitem[Huser et~al., 2016]{huser2016likelihood}
Huser, R., Davison, A.~C., and Genton, M.~G. (2016).
\newblock Likelihood estimators for multivariate extremes.
\newblock {\em Extremes}, 19(1):79--103.

\bibitem[Huser et~al., 2019]{huser2019full}
Huser, R., Dombry, C., Ribatet, M., and Genton, M.~G. (2019).
\newblock Full likelihood inference for max-stable data.
\newblock {\em Stat}, 8(1):e218.

\bibitem[Jiang et~al., 2017]{jiang2017learning}
Jiang, B., Wu, T.-y., Zheng, C., and Wong, W.~H. (2017).
\newblock Learning summary statistic for approximate {B}ayesian computation via
  deep neural network.
\newblock {\em Statistica Sinica}, pages 1595--1618.

\bibitem[Juba and Le, 2019]{juba2019precision}
Juba, B. and Le, H.~S. (2019).
\newblock Precision-recall versus accuracy and the role of large data sets.
\newblock In {\em Proceedings of the AAAI Conference on Artificial
  Intelligence}, volume~33, pages 4039--4048.

\bibitem[Kabluchko et~al., 2009]{kabluchko2009stationary}
Kabluchko, Z., Schlather, M., De~Haan, L., et~al. (2009).
\newblock Stationary max-stable fields associated to negative definite
  functions.
\newblock {\em The Annals of Probability}, 37(5):2042--2065.

\bibitem[Kharin and Zwiers, 2000]{kharin2000changes}
Kharin, V.~V. and Zwiers, F.~W. (2000).
\newblock Changes in the extremes in an ensemble of transient climate
  simulations with a coupled atmosphere--ocean gcm.
\newblock {\em Journal of Climate}, 13(21):3760--3788.

\bibitem[Kingma and Ba, 2014]{kingma2014adam}
Kingma, D.~P. and Ba, J. (2014).
\newblock Adam: A method for stochastic optimization.
\newblock {\em arXiv preprint arXiv:1412.6980}.

\bibitem[Lawson, 2008]{lawson2008bayesian}
Lawson, A.~B. (2008).
\newblock {\em Bayesian disease mapping: hierarchical modeling in spatial
  epidemiology}.
\newblock Chapman and Hall/CRC.

\bibitem[Liu et~al., 2018]{liu2018deep}
Liu, K., Ok, K., Vega-Brown, W., and Roy, N. (2018).
\newblock Deep inference for covariance estimation: Learning {G}aussian noise
  models for state estimation.
\newblock In {\em 2018 IEEE International Conference on Robotics and Automation
  (ICRA)}, pages 1436--1443. IEEE.

\bibitem[Mitchell et~al., 2004]{mitchell2004multi}
Mitchell, K.~E., Lohmann, D., Houser, P.~R., Wood, E.~F., Schaake, J.~C.,
  Robock, A., Cosgrove, B.~A., Sheffield, J., Duan, Q., Luo, L., et~al. (2004).
\newblock The multi-institution north american land data assimilation system
  ({N}{L}{D}{A}{S}): Utilizing multiple {G}{C}{I}{P} products and partners in a
  continental distributed hydrological modeling system.
\newblock {\em Journal of Geophysical Research: Atmospheres}, 109(D7).

\bibitem[Moller and Waagepetersen, 2003]{moller2003statistical}
Moller, J. and Waagepetersen, R.~P. (2003).
\newblock {\em Statistical inference and simulation for spatial point
  processes}.
\newblock CRC Press.

\bibitem[Padoan et~al., 2010]{padoan2010likelihood}
Padoan, S.~A., Ribatet, M., and Sisson, S.~A. (2010).
\newblock Likelihood-based inference for max-stable processes.
\newblock {\em Journal of the American Statistical Association},
  105(489):263--277.

\bibitem[Paszke et~al., 2017]{paszke2017automatic}
Paszke, A., Gross, S., Chintala, S., Chanan, G., Yang, E., DeVito, Z., Lin, Z.,
  Desmaison, A., Antiga, L., and Lerer, A. (2017).
\newblock Automatic differentiation in pytorch.

\bibitem[Pinaya et~al., 2020]{pinaya2020convolutional}
Pinaya, W. H.~L., Vieira, S., Garcia-Dias, R., and Mechelli, A. (2020).
\newblock Convolutional neural networks.
\newblock In {\em Machine Learning}, pages 173--191. Elsevier.

\bibitem[Radev et~al., 2020]{radev2020bayesflow}
Radev, S.~T., Mertens, U.~K., Voss, A., Ardizzone, L., and K{\"o}the, U.
  (2020).
\newblock Bayesflow: Learning complex stochastic models with invertible neural
  networks.
\newblock {\em IEEE Transactions on Neural Networks and Learning Systems}.

\bibitem[Ribatet, 2008]{ribatet2008spatialextremes}
Ribatet, M. (2008).
\newblock Spatial{E}xtremes: An {R}-package for modelling spatial extremes.
\newblock {\em R Package Version}, pages 2--0.

\bibitem[Ribatet et~al., 2012]{ribatet2012bayesian}
Ribatet, M., Cooley, D., and Davison, A.~C. (2012).
\newblock Bayesian inference from composite likelihoods, with an application to
  spatial extremes.
\newblock {\em Statistica Sinica}, pages 813--845.

\bibitem[Rudi et~al., 2021]{rudi2020parameter}
Rudi, J., Bessac, J., and Lenzi, A. (2021).
\newblock Parameter estimation with dense and convolutional neural networks
  applied to the {F}itzhugh-{N}agumo {O}{D}{E}.
\newblock {\em Mathematical and Scientific Machine Learning}.

\bibitem[Schlather, 2002]{schlather2002models}
Schlather, M. (2002).
\newblock Models for stationary max-stable random fields.
\newblock {\em Extremes}, 5(1):33--44.

\bibitem[Shang et~al., 2015]{shang2015two}
Shang, H., Yan, J., and Zhang, X. (2015).
\newblock A two-step approach to model precipitation extremes in {C}alifornia
  based on max-stable and marginal point processes.
\newblock {\em The Annals of Applied Statistics}, pages 452--473.

\bibitem[Stephenson and Tawn, 2005]{stephenson2005exploiting}
Stephenson, A. and Tawn, J. (2005).
\newblock Exploiting occurrence times in likelihood inference for componentwise
  maxima.
\newblock {\em Biometrika}, 92(1):213--227.

\bibitem[Tawn, 1988]{tawn1988bivariate}
Tawn, J.~A. (1988).
\newblock Bivariate extreme value theory: models and estimation.
\newblock {\em Biometrika}, 75(3):397--415.

\bibitem[Thibaud et~al., 2016]{thibaud2016bayesian}
Thibaud, E., Aalto, J., Cooley, D.~S., Davison, A.~C., Heikkinen, J., et~al.
  (2016).
\newblock Bayesian inference for the {B}rown--{R}esnick process, with an
  application to extreme low temperatures.
\newblock {\em The Annals of Applied Statistics}, 10(4):2303--2324.

\bibitem[Thibaud and Opitz, 2015]{thibaud2015efficient}
Thibaud, E. and Opitz, T. (2015).
\newblock Efficient inference and simulation for elliptical {P}areto processes.
\newblock {\em Biometrika}, 102(4):855--870.

\bibitem[Varin et~al., 2011]{varin2011overview}
Varin, C., Reid, N., and Firth, D. (2011).
\newblock An overview of composite likelihood methods.
\newblock {\em Statistica Sinica}, pages 5--42.

\bibitem[Wood, 2010]{wood2010statistical}
Wood, S.~N. (2010).
\newblock Statistical inference for noisy nonlinear ecological dynamic systems.
\newblock {\em Nature}, 466(7310):1102--1104.

\bibitem[Xia et~al., 2012a]{xia2012part2}
Xia, Y., Mitchell, K., Ek, M., Cosgrove, B., Sheffield, J., Luo, L., Alonge,
  C., Wei, H., Meng, J., Livneh, B., et~al. (2012a).
\newblock Continental-scale water and energy flux analysis and validation for
  north american land data assimilation system project phase 2 ({NLDAS}-2): 2.
  validation of model-simulated streamflow.
\newblock {\em Journal of Geophysical Research: Atmospheres}, 117(D3).

\bibitem[Xia et~al., 2012b]{xia2012part1}
Xia, Y., Mitchell, K., Ek, M., Sheffield, J., Cosgrove, B., Wood, E., Luo, L.,
  Alonge, C., Wei, H., Meng, J., et~al. (2012b).
\newblock Continental-scale water and energy flux analysis and validation for
  the north american land data assimilation system project phase 2 ({NLDAS}-2):
  1. intercomparison and application of model products.
\newblock {\em Journal of Geophysical Research: Atmospheres}, 117(D3).

\bibitem[Xu and Reid, 2011]{xu2011robustness}
Xu, X. and Reid, N. (2011).
\newblock On the robustness of maximum composite likelihood estimate.
\newblock {\em Journal of Statistical Planning and Inference},
  141(9):3047--3054.

\end{thebibliography}

\bibliographystyle{apalike}

\end{document}